\documentclass[aps,pra,eqsecnum,twocolumn]{revtex4}
\usepackage{graphicx}
\usepackage{bm}
\usepackage{amsmath,amssymb,amsthm,dsfont,bm,enumitem}
\usepackage{color}
\usepackage{wasysym}
\usepackage{ulem}
\usepackage[dvipsnames]{xcolor}

\usepackage[applemac]{inputenc}

\usepackage[colorlinks=true,urlcolor=blue,citecolor=blue,linkcolor=blue]{hyperref}

\begin{document}
\preprint{}
\title{Statistical analysis of Bell tests via generalized measurements}
\author{Alfredo Luis}
\email{alluis@fis.ucm.es}
\homepage{https://sites.google.com/ucm.es/alfredo/inicio}
\affiliation{Departamento de \'{O}ptica, Facultad de Ciencias
F\'{\i}sicas, Universidad Complutense, 28040 Madrid, Spain}
\date{\today}

\begin{abstract}
We provide a fully statistical analysis of the results of a Bell test beyond mean values. This is possible in a practical scheme where all the observables involved in the test are simultaneously measured at the expense of unavoidably additional noise. To deal with this noise we can follow two strategies leading to the same results. These are to adapt the Bell bound to include the noise, or to remove the additional noise via suitable data inversion. 
\end{abstract}

\maketitle

\section{Introduction}

Bell-like tests provide a powerful tool for investigating fundamental concepts at the very heart of the quantum theory \cite{LB90,JB64,WW,AF82,AR15,BKO16,CH74,CHSH69,MA84,MC}. Deep down, Bell tests are of profound statistical nature \cite{SGH21,SH22,AA24}, and so they should be a matter of statistical evaluation. Actually, the derivation of the Bell-type bounds demands a unique probability distribution of hidden variables to cover all experimental realizations at once, otherwise no meaningful bound might be derived \cite{MA84,MC,AK00,HP04,AM08,TN11,AK14,JCh17,NV62}.

However, a key feature of such tests is that they unavoidable involve the measurement of incompatible observables at each party. Because of this, the standard approach  requires to combine results from different incompatible experimental realizations, different physical contexts at different times. This prevents their joint statistical treatment beyond mean values. A deeper statistical evaluation is not allowed in standard approach precisely because of the lack of a unique common statistical framework. 

This serious difficulties can be avoided by addressing practical schemes where all the observables involved in the Bell test are determined, one way or the other, from a single experimental arrangement, where all measurements required are performed at once. This can be done via a generalized noisy joint measurement as already addressed in Refs. \cite{MM89,WMM02,MA84,MAL20,MAL22,VRA23}. The key point is to design the joint measurement such that its statistics provides enough information about all the observables involved in the corresponding Bell test. So, if necessary, their exact statistics can be obtained via a simple data inversion from the recorded data \cite{MM89,WMM02,MA84,WMM14,PB87, AL16,LM17,GBAL18,YLLO10}.

This offers a plenty of possibilities regarding practical and interpretational issues \cite{GP25}. For example this includes applying Bell tests to single measurements \cite{AHQ20,AL25}. Following this idea, in this work we provide a complete statistical evaluation of a Bell test within this generalized measurement scheme. This has been already carried out in Refs. \cite{VRA23,GP25} following an slightly different strategy. Within the framework addressed here we can readily evaluate the probability that an state fulfilling the test at the level of mean values may violate the test after a given finite number of trials, and vice versa. This question was left open for in Ref. \cite{AA24}, which is now possible to answer with the tools provided by generalized measurements.

To this end we will follow to equivalent routes to deal with the additional noise unavoidably introduced to make room for a the joint measurement. On the one hand, in Sec. III we derive a Bell inequality to be of application in a noisy environment. On the other hand, in Sec. IV we carry out the statistical analysis after noise removal. We will explore both ways showing their complete equivalence. 
 
\bigskip

\section{Settings}

Let us present the Bell test to be examined along with the basis for the joint measurement proposed for all the observables involved.

\subsection{CHSH Bell test}

We consider two subsystems, where two dichotomic observables can be measured in each subsystem, say $X$ and $Y$ in one of them and $U$, $V$ at the other. In these conditions it has been shown that in a classical-like scenario there is an upper bound to the modulus of the following combination of pairwise inter-subsystem correlations
\begin{equation}
    S= \langle XU \rangle - \langle XV \rangle + \langle YU \rangle + \langle YV \rangle  ,
\end{equation}
the bound being 
\begin{equation}
\label{ebS}
    |S | \leq 2 .
\end{equation}
What is exactly meant by classical-like scenario depends on some hypothesis regarding the statistical nature of the random variables considered. That is to say, the probability space where the test is planned, as suitable discussed in Ref. \cite{SGH21}.

\bigskip

\subsection{Noisy joint measurement}

We follow the general scheme in Ref. \cite{MAL20} for a noisy joint measurement of all the four dichotomic observables $X, Y, U, V$ involved in the Bell test. This is to be carried out via the exact, simultaneous, noise-free joint measurement of some compatible observables $X^\prime, Y^\prime, U^\prime, V^\prime$, with a four-tuple of dichotomic outcomes $\xi^\prime = (x^\prime,y^\prime,u^\prime,v^\prime)$. These are directly related to the original ones  $X, Y, U, V$ in a way to be precised soon. In quantum optics this seems particularly simple via suitable beam splitting for example. In each run of the experiment we get one outcome $\xi^\prime$ with a probability $p^\prime  (\xi^\prime | \rho )$ depending on the system state $\rho$ and the particular observation arrangement considered. 

Since this is a noisy simultaneous measurement we look for the general relation between noisy and noiseless statistics. For dichotomic variables there is a simple relation between the noisy marginals $p_{K^\prime}  (\kappa^\prime)$ and the noiseless ones $p_K (\kappa)$, for $K=X,Y,U,V$. Taking into account a natural unbiasedness requirement, regarding that if  $p_K (\kappa)$ is uniform then so is $p^\prime_{K^\prime} (\kappa^\prime)$ \cite{YLLO10}, and that $\kappa,\kappa^\prime = \pm 1$, this relation must be of the form
\begin{equation}
\label{nn1}
p^\prime_{K^\prime} (\kappa^\prime |\rho) = \sum_\kappa  p_K (\kappa^\prime|\kappa)  p_K (\kappa |\rho) , 
\end{equation}
where the conditional probabilities $p_K (\kappa^\prime|\kappa)$ are
\begin{equation}
\label{nn2}
p_K (\kappa^\prime|\kappa) = \frac{1}{2}  \left  ( 1 +  \gamma_K \kappa \kappa^\prime  \right ) , 
\end{equation}
and $\gamma_K$ are real factors with $|\gamma_K| \leq 1$. These factors are expressing the accuracy in the observation so that the mean value of each $K^\prime$
\begin{equation}
    \langle K^\prime \rangle = \sum_{\kappa^\prime} \kappa^\prime p^\prime_{K^\prime} (\kappa^\prime | \rho)  ,
\end{equation}
is related to the exact one
\begin{equation}
    \langle K \rangle = \sum_{\kappa} \kappa  p_K (\kappa | \rho),
\end{equation}
in the form 
\begin{equation}
\label{nmv}
\langle K^\prime \rangle = \gamma_K \langle K \rangle .
\end{equation}
Note that for dichotomic observables the variance is determined by the mean value
\begin{equation}
\Delta^2 K^\prime = 1 - \langle K^\prime \rangle^2, 
\end{equation}
so, since $|\langle K^\prime \rangle |\leq |\langle K \rangle |$ we have that $\Delta^2 K^\prime$ is always larger than $\Delta^2 K$ , the lesser the $|\gamma_K |$ the larger the difference. 

\bigskip

The factors $\gamma_K$ are state-independent being determined just by the particular measurement arrangement considered. They are not free parameters since they are constrained so that the joint $ p^\prime  (\xi^\prime | \rho) $ satisfies all necessary requirements to be a {\it bona fide} probability distribution, including positive semidefiniteness.  In some simple but meaningful realizations this means that \cite{MAL22}
\begin{equation}
    \gamma_X^2+ \gamma_Y^2 \leq 1, \qquad \gamma_U^2+ \gamma_V^2 \leq 1 .
\end{equation}
For simplicity we may consider the same amount of noise for all the observables, so that the gamma factors are all equal 
\begin{equation}
\label{ae}
    \gamma_X = \gamma_Y =\gamma_U = \gamma_V = \gamma.
\end{equation}

Let us note that all these relations between exact a noisy observables work equally well in the classical and quantum theories. They can be simply determined experimentally by procedures that in principle are independent of the theory.

\bigskip

\section{Noisy Bell test and its statistics}

In order to address the statistical evaluation of the Bell test we can follow two procedures that later we will show are completely equivalent. A first  simple possibility is to adapt the Bell test to a noisy environment with the derivation of new bound. Alternatively, in the next section we may carry out a noise removal via a suitable inversion procedure.

\subsection{Noisy Bell test}
Let us consider the adaptation of the Bell inequalities to a noisy scheme. To this end we consider a $s(\xi^\prime)$ random variable defined in terms of the outputs of the noisy joint measurement, that is
\begin{equation}
    s^\prime = s(\xi^\prime) = x^\prime u^\prime - x^\prime v^\prime + y^\prime u^\prime + y^\prime v^\prime  .
\end{equation}
This random variable con only take two values $s^\prime=\pm 2$ so in principle the same bound (\ref{ebS}) holds for $|S^\prime |$ as for $|S|$, say $|S^\prime | \leq 2$, where $S^\prime$ is the mean value of $s^\prime$
\begin{equation}
    S^\prime = \langle s^\prime \rangle = \sum_{\xi^\prime} s (\xi^\prime ) p^\prime  (\xi^\prime |\rho )  .
\end{equation}
However, since the measurement is noisy there is actually a tighter bound to $|S^\prime |$. To find such bound we note that each pair, such as $X^\prime$ and $U^\prime$ for example, are measured independently in each subsystem, so relations (\ref{nn1}) and (\ref{nn2}) are valid for both outcomes  
\begin{equation}
\label{nn3}
p^\prime_{X^\prime, U^\prime} (x^\prime,u^\prime |\rho) = \sum_\kappa  p_X (x^\prime|x) p_U (u^\prime|u)  p_{X,U} (x,u|\rho) , 
\end{equation}
and then
\begin{equation}
\label{SpS}
   S^\prime = \gamma^2 S .
\end{equation}
Finally, after Eq. (\ref{ebS}) we get a suitable form of Bell inequality adapted to this framework as 
\begin{equation}
\label{nbSp}
    |S^\prime | \leq  2 \gamma^2 .
\end{equation}
The violation of this bound is fully equivalent to the violation of the bound (\ref{ebS}).

\bigskip

\subsection{Statistics of the Bell test and its violation}

At difference with the original formulation of the bound (\ref{ebS}), in this case all the variables in $\xi^\prime$ are measured simultaneously in one and the same experimental arrangement. Thus we have access to a full statistics $p (s^\prime|\rho)$ for the $s^\prime$ variable beyond the mean value $S^\prime$. That is 
\begin{equation}
p (s^\prime |\rho) = \sum_{\xi^\prime} \delta_{s^\prime, s(\xi^\prime)} p^\prime  (\xi^\prime |\rho ) ,
\end{equation}
where $\delta$ is the Kronecker delta. Since there are only two possible values for $s^\prime$ we readily get
\begin{equation}
\label{fnsS}
p(s^\prime =\pm 2 |\rho) = \frac{1}{2} \pm \frac{S^\prime}{4} = \frac{1}{2} \pm \frac{\gamma^2 S}{4} .
\end{equation}
This is always a {\it bona fide} probability distribution $p(s^\prime =\pm 2 |\rho) \geq 0$. 

\bigskip

With this we can examine the probability that the result of a series of a finite number of $N$ measurements may violate the bound (\ref{nbSp}). In this regard, it is worth noting that for $N=1$ every single outcome $s^\prime = \pm 2$ violates the bound (\ref{nbSp}). This is in full accordance with the same conclusion obtained via noise-removing data inversion, shown in detail in Ref. \cite{AL25}.

Next we can examine this point as the number $N$ of repetitions of the measurement increases. To this end we begin with the probability of obtaining $n$ results $s^\prime= 2$ and $N-n$ results $s^\prime= -2$ after $N$ trials,
\begin{equation}
\label{pnN}
p(n|N)= \begin{pmatrix} N \cr n  \end{pmatrix} p^n(s^\prime=2|\rho )p^{N-n} (s^\prime =-2|\rho ) ,
\end{equation}
giving a mean value 
\begin{equation}
\label{mvSp}
S^\prime_N = 2 \left ( 2 \frac{n}{N} -1 \right ) .
\end{equation}
With this we can compute the probability $p(|S^\prime_N| > 2 \gamma^2)$ that after $N$ measurements the mean value $S^\prime_N$ violates the bound $2 \gamma^2$ in Eq. (\ref{nbSp}). After Eqs. (\ref{nbSp}) and (\ref{mvSp}) this violation holds for all results $n$ such that
\begin{equation}
\label{nviol}
    n > \frac{N}{2} \left (1+\gamma^2 \right ) , \quad n < \frac{N}{2} \left (1 -\gamma^2 \right )
\end{equation}
so the probability of violation of the bound is
\begin{equation}
\label{pvN}
   p(|S^\prime_N| > 2 \gamma^2 ) = 1- \sum_{n=n_c}^{n_f} p (n|N),
\end{equation}
where 
\begin{equation}
    n_f= \mathrm{ floor} \left [  \frac{N}{2} \left (1+\gamma^2 \right ) \right ], \quad n_c = \mathrm{ ceiling} \left [  \frac{N}{2} \left (1-\gamma^2 \right ) \right ].
\end{equation}
In Fig. 1 we have represented this probability of bound violation $ p(|S^\prime_N| > 2 \gamma^2 )$ as a function of $N$ for three cases, all them with $\gamma= 1/\sqrt{2}$. In black line there is the case $S=0$, which is well deep in the classical-like domain and provides no violation of the mean-value bound. In red line there is the fully quantum case $S = 2 \sqrt{2}$ providing maximum violation of the mean-value bound. Finally, in blue line we consider  the case $S=2$ providing no violation of the mean-value bound but being in the quantum-classical border. 

It can be appreciated that in the deep classical-like case $S=0$ in black line the violation probability tends to zero as $N$ increases, while in the extreme quantum case $S = 2 \sqrt{2}$ in red line tends to unity. On the other hand, in the border case $S=2$ in blue line it tends to a 50 \% of probability of obtaining violations, even in the limit of large $N$. 

In dashed lines we provide in Fig. 1 a suitable approximation derived after approximating  $p(n|N)$ by a Gaussian considering $n$ as a continuous variable, that is, 
\begin{equation}
p(n|N) \approx p_a(n|N) =\frac{\exp \left \{ - \frac{ \left [ n- (N/4) (2+ \gamma^2 S ) \right ]^2}{2 (N/16)\left (4-\gamma^4 S^2\right)}\right \} }{\sqrt{2 \pi (N/16)\left (4-\gamma^4 S^2\right)}} .
\end{equation}
This provides an approximate expression for the probability (\ref{pvN}) of bound violation of the form 
\begin{equation}
\label{apvN}
   p(|S^\prime_N| > 2 \gamma^2 ) \simeq  1- \int_{n_c}^{n_f} dn \; p_a (n|N) .
\end{equation}
We can see that the Gaussian approximation fits very well the exact computation even for small $N$. It allows us to understand the behavior of the violation probability. The integration in Eq. (\ref{apvN}) goes from  $N/2- N\gamma^2/2$ to $N/2 + N\gamma^2/2$, while the Gaussian is centered at $N/2+ N\gamma^2 S/4$. The width of the Gaussian scales as $\sqrt{N}$, so, in relative terms, it decreases as N increases. If $|S |<2$ the center of the Gaussian is within the integration limits, so for increasing $N$ the Gaussian will be increasingly contained within the integration range, and the violation probability will vanish. On the other hand, if $|S |>2$ the center of the Gaussian outside the integration limits, so for increasing $N$ the  Gaussian will be increasingly outside the integration range and the violation probability will tend to one. Finally, for $S=2$ half the Gaussian will be always within the integration range, being the other half always outside, leading to a violation of the test half of the times, no matter large $N$. 

\begin{figure}
\centering
\includegraphics[width=8cm]{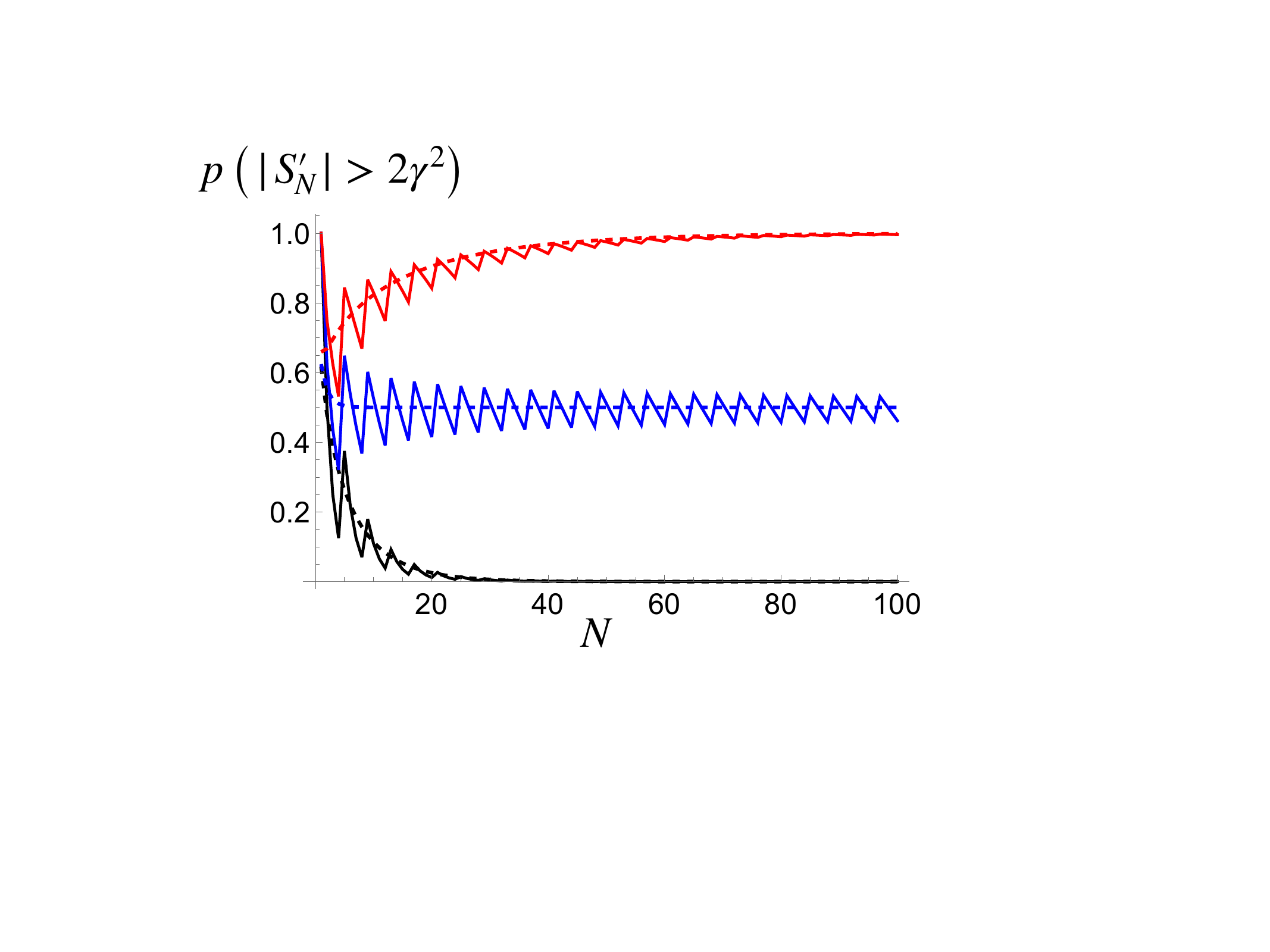}
\label{}
\caption{Probability of bound violation $ p(|S^\prime_N| > 2 \gamma^2 )$ as a function of $N$ for $S=0$ (black line), $S = 2 \sqrt{2}$ (red line), and $S=2$ (blue line), all them with $\gamma= 1/\sqrt{2}$. Dashed lines correspond to the Gaussian approximations in Eq. (\ref{apvN}). }
\end{figure}

\bigskip

\section{Equivalent bound violation after data inversion}

A relevant feature of the schemes we are considering is that the noisy can be easily removed, this is that relations  (\ref{nn1}), (\ref{nn2}) can be inverted to obtain the exact statistics $p_K (\kappa|\rho )$ in terms of the noisy ones $p_{K^\prime} (\kappa^\prime |\rho)$. That is that for each $K$ there are state-independent functions $\tilde{p}_K (\kappa | \kappa^\prime)$ such that
\begin{equation}
\label{pp1}
p_K (\kappa|\rho ) = \sum_{\kappa^\prime} \tilde{p}_K(\kappa |\kappa^\prime) p^\prime_{K^\prime} (\kappa^\prime |\rho) .
\end{equation}
This inverted distribution must satisfy
\begin{equation}
\sum_{\kappa^\prime} \tilde{p}_K(\kappa_1 |\kappa^\prime) p_K (\kappa^\prime|\kappa_2 ) = \delta_{\kappa_1,\kappa_2} = \frac{1}{2} \left ( 1 + \kappa_1 \kappa_2 \right ) ,
\end{equation}
where the last equality holds for dichotomic variables $\kappa = \pm 1$. After Eq. (\ref{nn2}) we readily get
\begin{equation}
\label{pp2}
p_K (\kappa | \kappa^\prime) = \frac{1}{2} \left ( 1 + \frac{\kappa \kappa^\prime}{ \gamma_K}  \right ) .
\end{equation}
We can collect all these inversions in a single joint distribution $p (\xi |\rho )$, where $\xi = ( x,y,u,v)$, defined as
\begin{equation}
\label{chr}
    p (\xi |\rho ) = \sum_{\xi^\prime} p (\xi |\xi^\prime ) p^\prime  ( \xi^\prime |\rho ) ,
\end{equation}
where
\begin{equation}
\label{pchichiprime}
    p (\xi |\xi^\prime )  = p_X (x | x^\prime )p_Y (y | y^\prime )p_U (u | u^\prime )p_V (v | v^\prime ) .
\end{equation}

We have shown in Ref. \cite{MAL20} that the violation of Bell inequality is equivalent to $p(\xi |\rho)$ taking negative values. Here we express this result in terms of the lack of a {\it bona fide} distribution for the random variable $s$
\begin{equation}
\label{sxi}
    s=s(\xi) = x u - x v + y u + y v .
\end{equation}
Such distribution can be constructed as
\begin{equation}
p(s |\rho) = \sum_\xi \delta_{s, s(\xi)} p(\xi|\rho ) ,
\end{equation}
where $ \delta_{s, s(\xi)}$ is the corresponding Kronecker delta. This naturally leads to  
\begin{equation}
p(s =\pm 2 |\rho) = \frac{1}{2} \pm \frac{S}{4} ,
\end{equation}
and violation of the Bell inequality, either (\ref{ebS}) or ( (\ref{nbSp}) is equivalent to $p(s|\rho)$ taking negative values. 

Because of these negative values, there is no way of  addressing of the statistics of the $s$ variable. A way round this difficulty arises if instead we consider a different strategy of data inversion. This is done after Eqs. (\ref{nmv}) and (\ref{SpS}) if we consider as random variable not $s$, but $s^\prime/\gamma^2$, given that the distribution $p(s^\prime|\rho)$ is well behaved. This is tantamount to consider as random variable the mean value of $s(\xi)$ in the one-shot noiseless inverted distribution $p(\xi|\xi^\prime)$ in Eq. (\ref{pchichiprime}). 

Here once again every outcome $s^\prime$ violates the noiseless bound (\ref{ebS}), in accordance with the negative values in $p(\xi |\xi^\prime)$. Regarding the trend to mean-value violation as the number $N$ of repetitions of the measurement increases we may follow a parallel analysis to the one already carried out above, with the very same probability $p(n|N)$ in Eq. (\ref{pnN}) of obtaining $n$ results $s = 2/\gamma^2$ and $N-n$ results $s= -2/\gamma^2$ after $N$ trials, to give giving a mean value 
\begin{equation}
S_N = \frac{2}{\gamma^2} \left ( 2 \frac{n}{N} -1 \right ) .
\end{equation}
With this we can compute the probability $p(|S_N| > 2)$ that after $N$ measurements this violates the bound Eq. (\ref{ebS}). This violation holds for exactly the same results $n$ in Eq. (\ref{nviol}) in the noisy version above, leading finally to the same results for the probability. 
This easily proves the equivalence between both formulations of the Bell test. 

\bigskip

\section{Conclusions}

We have provided a suitable statistical analysis of a Bell test within the framework of generalized measurements. Results have proven to be meaningful and consistent with the usual analysis in terms of mean values. Maybe the most interesting property of this analysis is being free of the inconveniences and difficulties caused by the unavoidable involvement of incompatible observables. We think this kind of analysis opens new routes worth following regarding the investigation of the fundamental physics involved in Bell-like tests.

\bigskip



\begin{thebibliography}{00}

\bibitem{LB90}
L. E. Ballentine, {\it Quantum Mechanics} (Prentice Hall, Englewood Cliffs, 1990). Chapter 20. 

\bibitem{JB64}
J. S. Bell, On the Einstein Podolsky Rosen paradox, \href{https://doi.org/10.1103/PhysicsPhysiqueFizika.1.195}{Physics {\bf 1}, 195--200 (1964)}.

\bibitem{WW}
R. F. Werner and M. M. Wolf, Bell inequalities and Entanglement, 
\href{https://doi.org/10.48550/arXiv.quant-ph/0107093}{arXiv:quant-ph/0107093}.

\bibitem{AF82}
A. Fine, Hidden Variables, Joint Probability, and the Bell Inequalities, \href{https://doi.org/10.1103/PhysRevLett.48.291}{Phys. Rev. Lett {\bf 48}, 291--295 (1982)}.

\bibitem{AR15}
A. Rivas, On the role of joint probability distributions of incompatible observables in Bell and Kochen--Specker Theorems, \href{https://doi.org/10.1016/j.aop.2019.167939}{Ann. Phys. (N.Y.) {\bf 411}, 167939 (2019)}.

\bibitem{BKO16}
J. A. de Barros, J. V. Kujala and G. Oas, Negative probabilities and contextuality, \href{https://doi.org/10.1016/j.jmp.2016.04.014}{J. Math. Psychol.  {\bf 74},  34--45 (2016)}. 

\bibitem{CH74}
J.F. Clauser and M.A. Horne, Experimental consequences of objective local theories, \href{https://doi.org/10.1103/PhysRevD.10.526}{Phys. Rev. D {\bf 10}, 526--535 (1974)}.

\bibitem{CHSH69}
J. F. Clauser, M. A. Horne, A. Shimony, and R. A.  Holt, Proposed experiment to test local hidden-variable theories, \href{https://doi.org/10.1103/PhysRevLett.23.880}{Phys. Rev. Lett. {\bf 23}, 880--884 (1969)}.

\bibitem{MA84}
W. M. Muynck and O. Abu-Zeid, On an alternative interpretation of the Bell inequalities, \href{https://doi.org/10.1016/0375-9601(84)90832-6}{Phys. Lett. A {\bf 100}, 485--489 (1984)}.

\bibitem{MC}
M. Czachor, On some class of random variables leading to violations of the Bell inequality, \href{https://doi.org/10.1016/0375-9601(88)90334-9}{Phys. Lett. A {\bf 129}, 291--294 (1988)}; Erratum, \href{https://doi.org/10.1016/0375-9601(89)90697-X}{Phys. Lett. A {\bf 134}, 512(E) (1989)}.

\bibitem{SGH21}
A. F. G. Solis-Labastida, M. Gastelum, and J. G. Hirsch, The Violation of Bell-CHSH Inequalities Leads to Different Conclusions Depending on the Description Used, \href{https://doi.org/10.3390/e23070872}{Entropy  {\bf 23}, 872 (2021)}.

\bibitem{SH22}
A. F. G. Solis-Labastida and J. G. Hirsch, 
Remarks on the use of objective probabilities in Bell-CHSH inequalities, \href{https://doi.org/10.48550/arXiv.2202.08353}{arXiv:2202.08353v1 [quant-ph]} .

\bibitem{AA24}
A. Aiello, A possible statistics loophole in Bell's theorem, \href{https://doi.org/10.48550/arXiv.2412.17857}{arXiv:2412.17857v3 [physics.gen-ph]}.

\bibitem{AK00}
A. Khrennikov, Non-Kolmogorov probability models and modified Bell's inequality, \href{https://doi.org/10.1063/1.533210}{J. Math. Phys. {\bf 41}, 1768--1777 (2000)}.

\bibitem{HP04}
K. Hess and W. Philipp, Bell's theorem: Critique of proofs with and without inequalities, \href{https://doi.org/10.1063/1.1874568}{AIP Conf. Proc. {\bf 750}, 150--157 (2005)}.

\bibitem{AM08}
A. Matzkin, Is Bell's theorem relevant to quantum mechanics. On locality and non-commuting observables, \href{https://doi.org/10.1063/1.3109959}{AIP Conf. Proc. {\bf 1101}, 339--348 (2009)}.

 \bibitem{TN11}
T. M. Nieuwenhuizen, Is the Contextuality Loophole Fatal for the Derivation of Bell Inequalities?, \href{https://doi.org/10.1007/s10701-010-9461-z}{Found. Phys. {\bf 41}, 580--591 (2011)}.

\bibitem{AK14}
A. Khrennikov, CHSH inequality: Quantum probabilities as classical conditional probabilities, \href{https://doi.org/10.1007/s10701-014-9851-8}{Found. Phys. {\bf 45}, 711--725 (2015)}.

\bibitem{JCh17}
J. Christian, On a Surprising Oversight by John S. Bell in the Proof of his Famous Theorem, \href{https://doi.org/10.48550/arXiv.1704.02876}{arXiv:1704.02876 [physics.gen-ph]}.

\bibitem{NV62}
N. N. Vorob'ev, Consistent families of measures and their extensions, \href{https://doi.org/10.1137/1107014}{Theor. Probab. Applications {\bf VII}, 147 (1962)}.



\bibitem{MM89} 
W. M. de Muynck and H. Martens, Joint measurement of incompatibles observables and the Bell inequalities, \href{https://doi.org/10.1016/0375-9601(89)90310-1}{Phys. Lett. A \textbf{142}, 187--190 (1989)}.

\bibitem{WMM02}
W. M. Muynck, \textit{Foundations of Quantum Mechanics, an Empiricist Approach},  (Kluwer Academic Publishers,  2002).

\bibitem{MAL20}
E. Masa, L. Ares, and A. Luis, Nonclassical joint distributions and Bell measurements, \href{https://doi.org/10.1016/j.physleta.2020.126416}{Phys. Lett. A {\bf 384}, 126416  (2020)}.

\bibitem{MAL22}
E. Masa, L. Ares, and A. Luis, Inequalities for complementarity in observed statistics, \href{https://doi.org/10.1016/j.physleta.2021.127914}{Phys. Lett. A {\bf 427}, 127914 (2022)}.

\bibitem{VRA23}
S. Virz\'i, E. Rebufello, F. Atzori, A. Avella, F. Piacentini, R. Lussana, I. Cusini, F. Madonini, F. Villa, M. Gramegna, E. Cohen, I. P. Degiovanni, and M. Genovese, Entanglement-preserving measurement of the Bell parameter on a single entangled pair, \href{https://doi.org/10.1088/2058-9565/ad6a37}{Quantum Sci. Technol. {\bf 9}, 045027 (2024)}.

\bibitem{WMM14}
W. M. Muynck, Interpretations of quantum mechanics, and interpretations of violation of Bell's inequality, \href{https://doi.org/10.1142/9789812810809_0007}{Foundations of Probability and Physics, 95--114 (2001)}.

\bibitem{PB87}
P. Busch, Some Realizable Joint Measurements of Complementary Observables, \href{https://doi.org/10.1007/BF00734320}{Found.Phys., {\bf 17}, 905--937 (1987)}.

\bibitem{AL16}
A. Luis, Nonclassical light revealed by the joint statistics of simultaneous measurements, \href{http://dx.doi.org/10.1364/OL.41.001789}{Opt. Lett. {\bf 41}, 1789--1792 (2016)}.

\bibitem{LM17}
A. Luis and L. Monroy, Nonclassicality of coherent states: Entanglement of joint statistics, \href{https://doi.org/10.1103/PhysRevA.96.063802}{Phys. Rev A {\bf 96}, 063802 (2017)}.

\bibitem{GBAL18}
R. Galazo, I. Bartolom\'e, L. Ares, and A. Luis, Classical and quantum complementarity, \href{https://doi.org/10.1016/j.physleta.2020.126849}{Physics Letters A {\bf 384}, 126849 (2020)}.


\bibitem{YLLO10}
S. Yu, N. Liu, L. Li, and C. H. Oh, Joint measurement of two unsharp observables of a qubit, \href{https://doi.org/10.1103/PhysRevA.81.062116}{Phys. Rev. A {\bf 81}, 062116 (2010)}.

\bibitem{GP25}
M. Genovese and F. Piacentini, Consequences of the single-pair measurement of the Bell parameter, \href{https://doi.org/10.1103/PhysRevA.111.022204}{Phys. Rev. A {\bf 111}022204 (2025)}.

\bibitem{AHQ20}
M. Ara\'{u}jo, F. Hirsch, and M. T. Quintino, Bell nonlocality with a single shot, 
\href{https://doi.org/10.22331/q-2020-10-28-353}{Quantum {\bf 4}, 353 (2020)} .

\bibitem{AL25}
A., Luis, Single-measurement Bell analysis, \href{ https://doi.org/10.1103/PhysRevA.111.012213}{Phys. Rev. A {\bf 111}, 012213 (2025)}.




\end{thebibliography}
\end{document}